\documentclass[manuscript]{acmart}

\AtBeginDocument{%
  }

\copyrightyear{2025}
\acmYear{2025}
\setcopyright{rightsretained}
\acmConference[CHI EA '25]{Extended Abstracts of the CHI Conference on Human Factors in Computing Systems}{April 26-May 1, 2025}{Yokohama, Japan}
\acmBooktitle{Extended Abstracts of the CHI Conference on Human Factors in Computing Systems (CHI EA '25), April 26-May 1, 2025, Yokohama, Japan}\acmDOI{10.1145/3706599.3716227}
\acmISBN{979-8-4007-1395-8/25/04}

\begin{document}

\title{XAIxArts Manifesto: Explainable AI for the Arts}


\author{Nick Bryan-Kinns} 
\affiliation{%
  \institution{University of the Arts London}
  \city{London}
  \country{UK}}
\email{n.bryankinns@arts.ac.uk}
\orcid{0000-0002-1382-2914}

\author{Shuoyang Jasper Zheng} 
\affiliation{%
  \institution{Queen Mary University of London}
  \city{London}
  \country{UK}}
\email{shuoyang.zheng@qmul.ac.uk}
\orcid{0000-0002-5483-6028}

\author{Francisco Castro} 
\affiliation{%
  \institution{New York University}
  \city{New York}
  \country{United States}}
\email{francisco.castro@nyu.edu}
\orcid{0000-0002-3940-5334}

\author{Makayla Lewis} 
\affiliation{%
  \institution{Kingston University}
  \city{London}
  \country{UK}}
\email{m.m.lewis@kingston.ac.uk}
\orcid{0000-0002-4429-3402}

\author{Jia-Rey Chang} 
\affiliation{%
  \institution{Queen's University Belfast}
  \city{Belfast}
  \country{UK}}
\email{j.chang@qub.ac.uk}
\orcid{0009-0004-0058-7769}

\author{Gabriel Vigliensoni} 
\affiliation{%
  \institution{Concordia University}
  \city{Montréal}
  \country{Canada}}
\email{gabriel.vigliensoni@concordia.ca}
\orcid{0000-0003-0274-4356}

\author{Terence Broad} 
\affiliation{%
  \institution{University of the Arts London}
  \city{London}
  \country{UK}}
\email{t.broad@arts.ac.uk}
\orcid{0000-0001-9987-6536}

\author{Michael Clemens} 
\affiliation{%
  \institution{University of Utah}
  \city{Salt Lake City}
  \country{United States}}
\email{michael.clemens@utah.edu}
\orcid{0000-0002-4507-8421}

\author{Elizabeth Wilson} 
\affiliation{
    \institution{University of the Arts London}
    \city{London}
    \country{UK}}
\email{e.j.wilson@arts.ac.uk}
\orcid{0000-0001-6212-6627}

\renewcommand{\shortauthors}{Bryan-Kinns et al.}

\newcommand{\notes}[1]{\textcolor[RGB]{0, 153, 1}{[\textit{#1}]}}

\newcommand{\rr}[1]{#1}

\begin{abstract}
Explainable AI (XAI) is concerned with how to make AI models more understandable to people. To date these explanations have predominantly been technocentric - mechanistic or productivity oriented. This paper introduces the Explainable AI for the Arts (XAIxArts) manifesto to provoke new ways of thinking about explainability and AI beyond technocentric discourses.
\rr{Manifestos offer a means to communicate ideas, amplify unheard voices, and foster reflection on practice.
To supports the co-creation and revision of the XAIxArts manifesto we combine a World Café style discussion format with a living manifesto to question four core themes: 1) Empowerment, Inclusion, and Fairness; 2) Valuing Artistic Practice; 3) Hacking and Glitches; and 4) Openness.
Through our interactive living manifesto experience we invite participants to actively engage in shaping this XIAxArts vision within the CHI community and beyond.}
\end{abstract}

\ccsdesc[500]{Human-centered computing~Human computer interaction (HCI)}
\ccsdesc[500]{Human-centered computing~Interaction design}
\ccsdesc[500]{Human-centered computing~Visualization}
\ccsdesc[500]{Applied computing~Arts and humanities}
\ccsdesc[500]{Computing methodologies~Artificial intelligence}

\keywords{Manifesto, Explainable AI (XAI), Artificial Intelligence (AI), Arts, Generative Arts, AI Arts, Digital Arts, Human-Centred AI, Human-Computer Interaction (HCI), Interaction Design}


\begin{teaserfigure}
  \centerline{\includegraphics[width=0.38\textwidth]{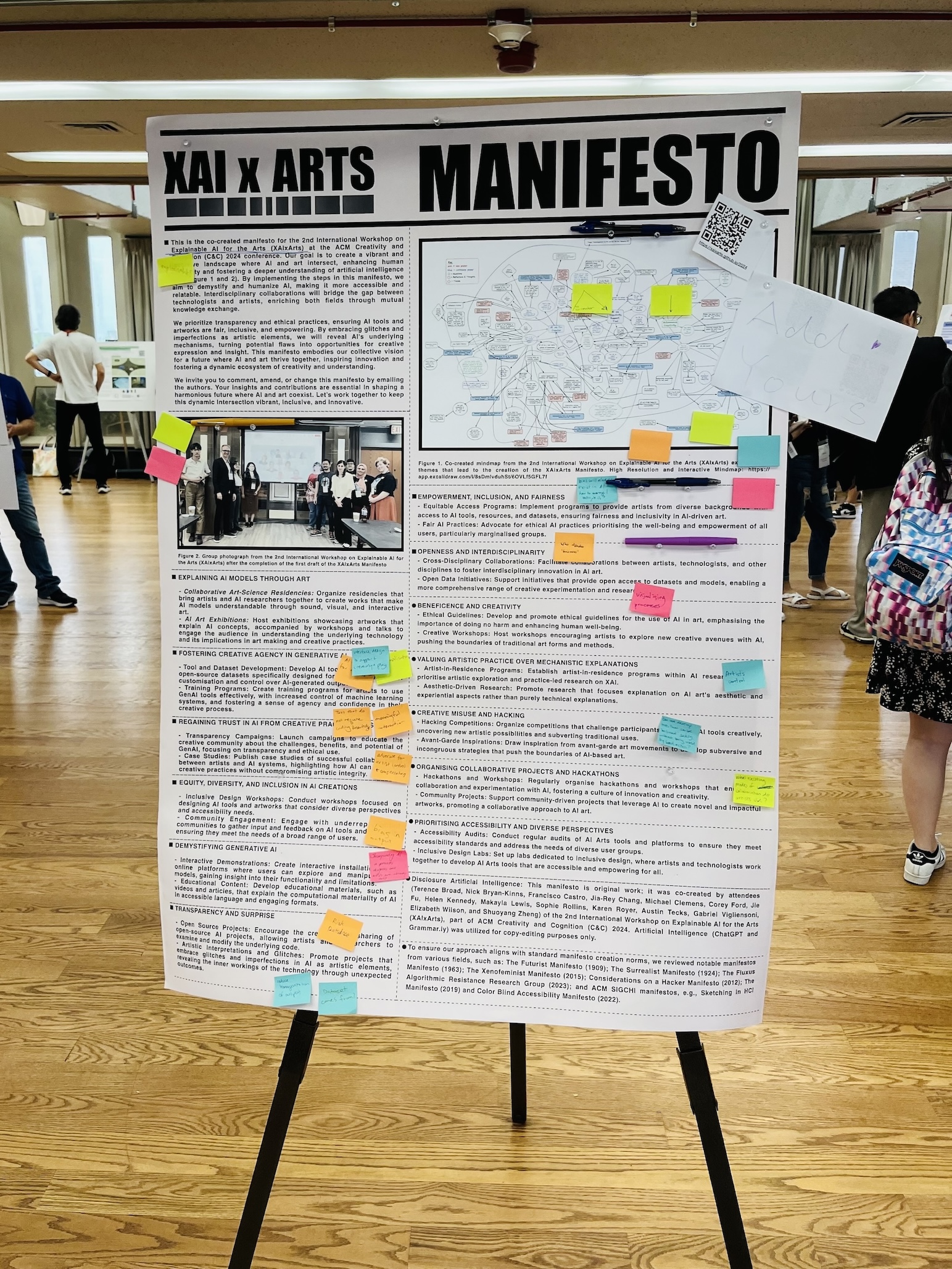}}
  \caption{Early version of the XAIxArts manifesto as a poster at the ACM Creativity and Cognition (C\&C) 2024 conference with post-it notes added by conference attendees.}
  \Description{A picture showing a printed out graphically designed poster of an early version of the XAIxArts manifesto. The poster is placed within the conference poster session room and has multicolored post-it notes attached to it by conference delegates to offer a chance to add feedback to the manifesto.}
  \label{fig:XAIxArtsPoster}
\end{teaserfigure}


\maketitle

\section{Introduction}

The field of Explainable AI (XAI) \cite{gunning_xaiexplainable_2019} is concerned with how to make AI models more understandable to people and has been a growing theme at Human-Computer Interaction (HCI) conferences including CHI, for example \citet{Zhang2022, Panigutti2022, Ehsan2022} and the ``Mistakes, Explainability'' session at the ACM CHI 2022 conference.
However, and somewhat ironically, explainability and explanation are ambiguously defined in current literature \cite{Ciatto2020}. For example, in machine learning (ML) literature, explainability usually refers to making the reasons behind ML decisions more comprehensible to people.
We subscribe to a broader view of the meaning of explainability in which ``explainability encompasses everything that makes ML models transparent and understandable, also including information about the data, performance, etc.'' \cite{Liao-2020}. From this point of view, explainability involves clarifying how AI models work and the reasoning behind their decisions, identifying and addressing biases in training datasets, understanding how these biases affect the models, and explaining the impact of AI models on energy consumption, the environment, and society.

XAI approaches are part of the broader fields of Responsible AI and Human-Centred AI \cite{shneiderman2022HCAI,garibay2023}, which have also been growing themes within HCI and the CHI community. Whilst XAI has been explored in areas such as healthcare \cite{quellec2021explain} and automation \cite{shen2020explain}, the explanations have predominantly been technocentric - mechanistic or productivity oriented - focusing on explaining functional aspects of AI models or the reasoning behind specific decisions made by AI systems as illustrated in the survey of \citet{guidotti2018survey}. In response to this, the XAIxArts workshop series was started in 2023 \cite{bryankinns2023proceedings} to investigate how XAI might be applied and understood within the Arts, and how the Arts could provide an ``alternative lens through which to examine XAI'' \cite{bryankinns2024proceedings}. Here the Arts inclusively refers to all artistic practices from music to visual arts, from dance to sculpture, or from poetry to film making.

Our XAIxArts international community has brought together over 60 academics and practising artists to date. The artistic practices in our community involve methods, techniques, and conceptual approaches used across creative disciplines of human expression, culture, creativity, and design. Works from our XAIxArts workshops can be found on the XAIxArts website\footnote{\url{https://xaixarts.github.io/}}. In our workshops we explored questions such as the role of XAI as an artistic material \cite{tecks_explainability_2024,broad_using_2024}, how Arts practices could be used to explain \cite{arandas_antagonising_2023} and navigate AI models \cite{wilson_embodied_2024}, and explaining artistic practices with AI \cite{lewis_aixartist_2023}. We explored how AI models could be made more explainable and controllable by artists \cite{clemens_explaining_2023,zheng_mapping_2024}, how bias in generative models could be better explained \cite{chang_loki_2024}, and how AI-generated errors could be captured and repurposed as part of artistic practice \cite{knight_artistic_2023}.
Through these explorations of XAI for the Arts, as a community, we identified challenges and opportunities for XAI through the lens of the Arts beyond the predominant technocentric discourses of explainability. Challenges include the lack of explainable AI contributing to barriers to equitable access to AI Arts tools by diverse user groups, and the atomised, isolated experiences that individual artists and creative practitioners are having with AI models. 
Opportunities include valuing artistic practice over mechanistic explanations of AI in order to offer alternative insights into AI, to demystify AI, and to build trust in AI. 

To move the field of XAIxArts forward, we propose constructive and actionable approaches in the form of a manifesto - a call to action approach \cite{hanna_dissent_2019} that advocates for change, to act rather than critique, and to concretize steps rather than dissect problems. Manifestos have historically served as powerful catalysts for innovation and progress across various domains, from art and design to technology and social movements \cite{marinetti2009futurist, breton_first_1924, wark2012considerations, monaco2022color}. Manifestos can effectively galvanize communities and drive positive change by presenting concrete steps and actionable items \cite{ashby_articulating_2023}. A manifesto gives voice to unheard views and scaffolds an ongoing process of reflective discourse. In the context of XAIxArts, this approach is important as it encourages developing and implementing solutions that address the unique challenges and opportunities at the intersection of Explainable AI and the Arts. Rather than solely critiquing existing limitations or speculating on potential futures, this manifesto aims to inspire action and facilitate tangible progress. It is a call to embrace experimentation, collaboration, and innovation, and to actively shape the future of XAIxArts. \rr{We aim to engage with a broad spectrum of audiences with our XAIxArts manifesto. Our core audience brings together artists, HCI academics including ``Artists as Researchers'' and ``Researchers as Artists'' \cite{sturdee_plurality_2021},  and technologists interested in AI and the Arts.}

We developed the first version of our XAIxArts manifesto in our 2024 workshop on XAIxArts \cite{bryankinns2024proceedings} to highlight these challenges and opportunities, and to drive positive human-centered change in Explainable AI for the Arts, and XAI more broadly. This paper marks the first publication of our XAIxArts manifesto and aims to engage a broader audience in our vision for a future where AI and art merge to foster innovation and understanding. It serves as a platform to disseminate the principles and actionable approaches to explainable AI for the Arts, inviting the broader community to contribute to its evolution and implementation. By extending the reach of the manifesto through this publication, we aim to spark further dialogue and collaboration, driving positive human-centered change in the field of XAIxArts.

\section{Manifesto Development}

\begin{figure*}
  \includegraphics[width=\textwidth]{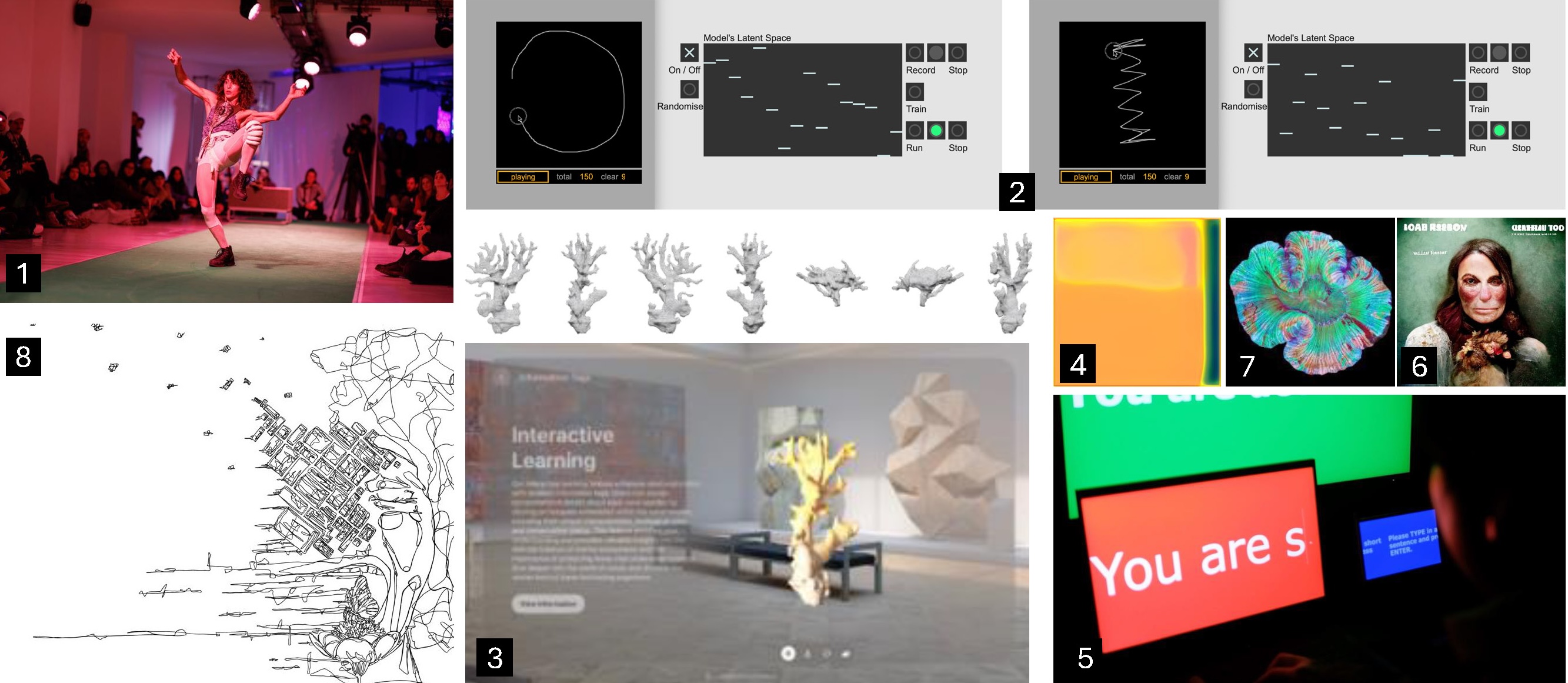}
  \caption{Authors Gallery from XAIxArts  Workshop 2 at Creativity and Cognition 2024: (1) Performer Manoela Rangel as part of the Patterns In Between Intelligences project explores latent decoding with sensors on the body \cite{wilson_embodied_2024}, Image Credit: Frank Sperling, \url{https://franksperling.net/Info};  (2) A sketching interface for neural audio synthesis models \cite{zheng_mapping_2024}, MaxMSP and nn\~{}, Shuoyang Zheng, Anna Xamb\'o Sed\'o and Nick Bryan-Kinns, 2024; (3) Generation Process of a Single Coral Model and Coral Database Interface Design on Vision Pro \cite{fu2024coralmodelgenerationsingle}; (4) The artwork (un)stable equilibrium by Terence Broad, 2019 \cite{broad_using_2024}; (5) Loki Test \cite{chang_loki_2024} \url{https://vimeo.com/904719832/d8b045a5b1}, TouchDesigner with ChatGPT API, Jia-Rey Chang, 2024; (6) The artwork LOAB by Steph Maj Swanson, 2022 \cite{broad_using_2024}; (7) 3D Coral Converted by AI Software and Processed into USDZ Format Using Zbrush \cite{fu2024coralmodelgenerationsingle}; and (8) Looking Back, Moving Forward was exhibited at "Spaces of Enquiry" at the Stanley Picker Gallery in the UK, Apple Pencil is on iPad Pro using the Procreate app, Makayla Lewis, 2024 \cite{lewis2024lookingbackmovingforward}.}
  \Description{(1) A woman wearing a white sensor body suit on her body, with one leg and two arms extended, in a gallery space with a crowd sitting and standing at the sides of the room. (2) Two screenshots of a digital sketching interface with 16 sliders as control parameters, one with a circle shape sketch, the other with a jagged shape sketch. (3) This is a 3D model generated from a single image using AI. Through an optimised workflow in 3D software, the model is imported into a VR environment while preserving and showcasing its vivid colour details. (4) An image generated by a neural network trained without any data. This is an abstract image with rough geometric blocks of colours, including orange, yellow, green and blue. (5) A screenshot of the "Loki Test" demonstrates how users engage with the AI chatbot by typing on a keyboard, like typical online messaging, and receiving sarcastic real-time feedback from Loki displayed on the screen. (6) AI-generated an image of a woman with brown hair on a green background with illegible text surrounding it. (7) This 3D model is transformed from a 2D coral photograph using AI technology. After detailed refinement in Zbrush, it accurately restores the intricate texture details of the original 2D image, delivering a more realistic visual experience. (8) Makayla screaming at a butterfly conducting its everyday activities and interacting with flowers coming from space in small to large fragments is a structured and scary block smacking into their head, Makayla's face appears stressed.}
  \label{fig:Gallery}
\end{figure*}

To co-create the first version of the XAIxArts manifesto, we brought together a community of 39 participants of researchers and creative practitioners in Human-Computer Interaction (HCI), Interaction Design, AI, explainable AI (XAI), and Digital Arts at the second International Workshop on Explainable AI for the Arts (XAIxArts) at the ACM Creativity and Cognition conference 2024. The goal of the workshop was to create a wide and inclusive community where AI and Arts practices intersect. Following a peer-review process, eleven submissions were accepted to be presented in a hybrid format at the workshop to collect insights and critically reflect on current and emerging practices in XAIxArts. Figure \ref{fig:Gallery} illustrates a selection of works shown by authors at the second XAIxArts workshop.

\begin{figure}
  \centerline{\includegraphics[width=0.75\columnwidth]{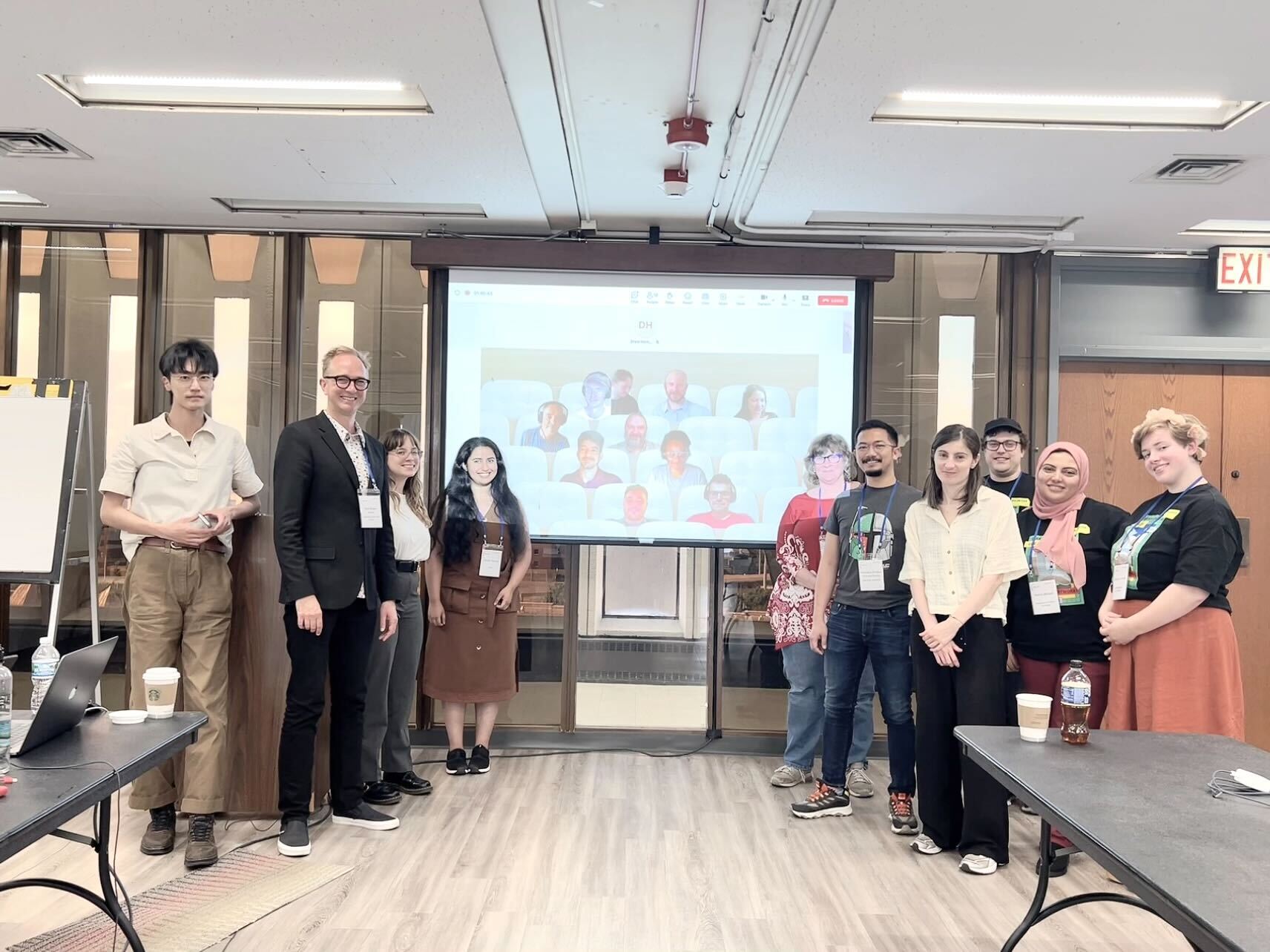}}
  \caption{Participants of the 2nd International Workshop on Explainable AI for the Arts (XAIxArts)
}
  \Description{Group photo showing participants at the 2nd International Workshop on Explainable AI for the Arts (XAIxArts) in person in Chicago, USA.}
  \label{fig:participants}
\end{figure}

\begin{figure*}
  \centerline{\includegraphics[width=1.0\textwidth]{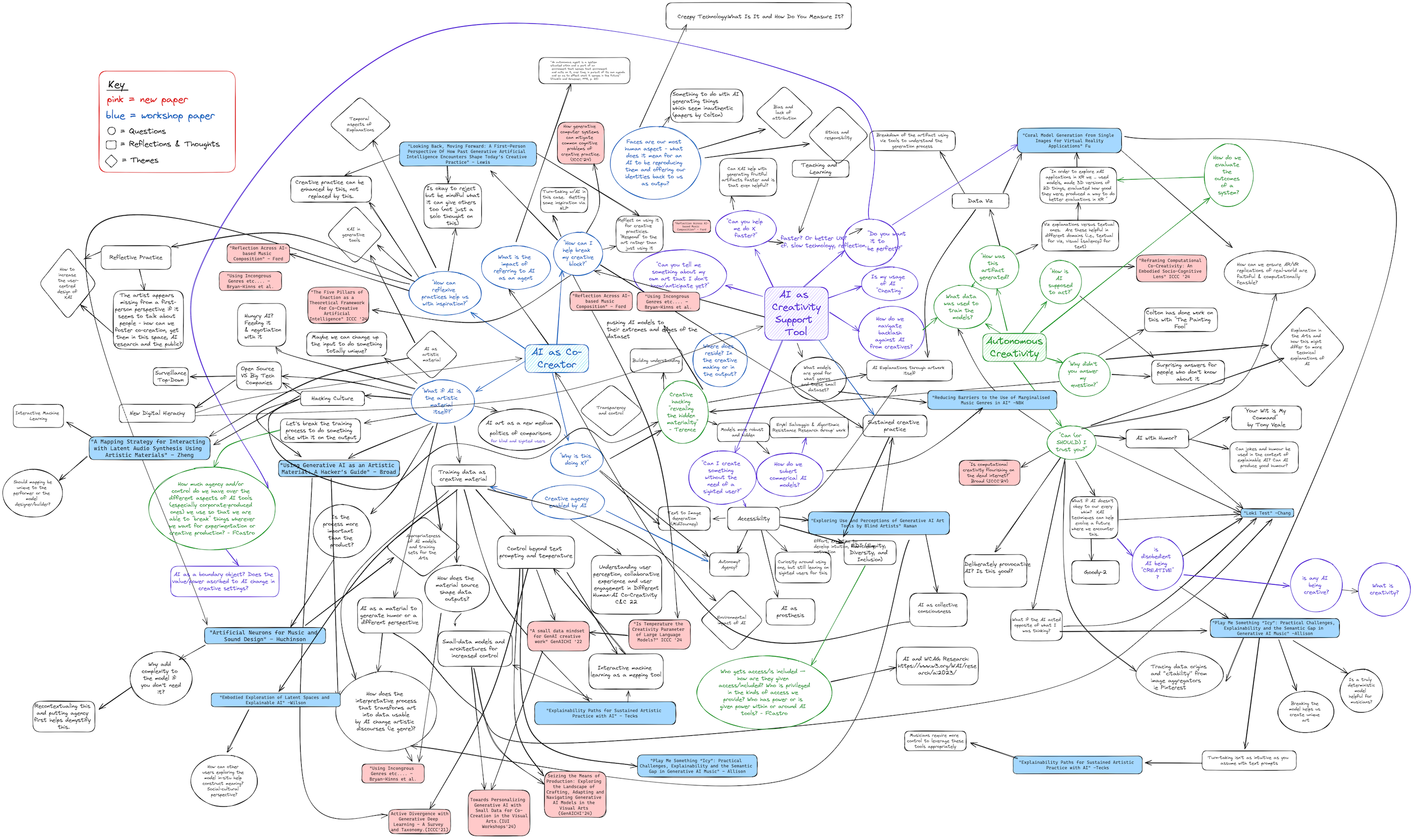}}
  \caption{Co-created mind-map from the 2nd International Workshop on Explainable AI for the Arts (XAIxArts) exploring the themes that lead to the creation of the XAIxArts Manifesto. High Resolution and Interactive Mindmap: \url{https://app.excalidraw.com/l/8sDmlvduhSt/6OVLf5GFL7f}
}
  \Description{Co-created mind-map exploring the themes that lead to the creation of the XAIxArts Manifesto.}
  \label{fig:mindmap}
\end{figure*}

To develop the first version of the XAIxArts manifesto, workshop participants (Figure \ref{fig:participants}) undertook two hybrid brainstorming activities to develop a mindmap (Figure~\ref{fig:mindmap}) that captured reflections, thoughts, and questions about the artistic and research work presented in the workshop (rectangles in the figure), and other work related to XAIxArts. 
The brainstorming activities were followed by each participant reflecting on the content they had contributed to the mindmap. The mindmap was then collaboratively organised around themes that emerged in the workshop (diamond-shapes in Figure~\ref{fig:mindmap}).
The XAIxArts Manifesto was then created to summarize key topics and themes identified in the mindmap. The authors drafted the manifesto over a period of two days after the workshop using a collaborative document editor. The final version of the XAIxArts manifesto was then drafted in poster format, which was printed and displayed at the ACM Creativity and Cognition 2024 conference. Conference attendees were invited to add comments and thoughts to the manifesto using post-it notes to the poster, as shown in Figure \ref{fig:XAIxArtsPoster}. These contributions were used to refine the text further, with a final collaborative iteration of editing the manifesto text taking place online to produce the XAIxArts manifesto presented in this paper for the first time.
\rr{Our XAIxArts manifesto complements established initiatives in AI and the Arts such as large scale-scale institutional practices of Future Art Ecosystems (FAE)\footnote{\url{https://futureartecosystems.org/}}. We see approaches such as FAE as focussed more on eco-system and infrastructure building and long-term strategy whereas our XAIxArts manifesto is a call for action for individuals to change our AI and Arts infrastructures and practices and to reframe our ways of thinking about AI and the Arts through an explainable AI lens. }

To ensure our approach aligns with standard manifesto creation norms, we reviewed notable manifestos from various fields including: The Futurist Manifesto (1909) \cite{marinetti2009futurist}; The Surrealist Manifesto (1924) \cite{breton_first_1924}; The Fluxus Manifesto (1963) \cite{maciunas_fluxus_1963}; The Xenofeminist Manifesto (2018) \cite{cuboniks_xenofeminist_2018}; Considerations on a Hacker Manifesto (2012) \cite{wark2012considerations}; The Algorithmic Resistance Research Group (2023) \cite{salvaggio_algorithmic_2023}; 
and ACM SIGCHI manifestos, e.g. Sketching in HCI Manifesto (2019) \cite{lewis_sketching_2019} and Color Blind Accessibility Manifesto (2022) \cite{monaco2022color}.

\section{XAIxArts Manifesto 1.0}\label{sec:manifesto}

\rr{This is our \emph{Explainable AI for the Arts (XAIxArts) Manifesto}.} It prioritises transparent and ethical practices, provoking us to ensure AI tools and artworks are fair, inclusive, and empowering. By embracing glitches and imperfections as artistic elements, we will reveal AI's underlying mechanisms, turning potential flaws into opportunities for creative expression and insight. This manifesto embodies our collective vision for a future where AI and the Arts thrive together, inspiring innovation and fostering a dynamic ecosystem of creativity and understanding beyond dominant technocentric concerns of productivity and efficiency. We believe that explainability is a key to unlocking this future.

\subsection*{Empowerment, Inclusion, and Fairness}

Explainable AI is fundamentally a human-centred approach to AI design. It offers opportunities for more transparent understanding of AI models and interaction with AI. However, poorly designed XAI or XAI focused purely on technocentric values may raise barriers to AI use and bias AI further towards metrics of productivity and efficiency, which do not positively contribute to human well-being. We ask you to:

\subsubsection*{Fair AI Practices} Advocate for ethical XAI practices recognising the inherent bias in AI \rr{\cite{arandas_antagonising_2023}}, prioritising the well-being and empowerment of all users, particularly marginalised groups, building a risk-register of inequity, and gaining consent for any data that is used in AI training and inference. 

\subsubsection*{Ethical Guidelines} Develop and promote ethical guidelines for the use of AI in art, emphasising the importance of explainability contributing to doing no harm and enhancing human well-being.

\subsubsection*{Community Engagement} Engage with underrepresented communities to gather input and feedback on AI tools, datasets, and their bias, to inform future XAI ensuring they meet the needs of a broad range of users.

\subsubsection*{Equitable Access Programs} Co-develop artist programs to provide artists from diverse backgrounds and with diverse abilities  access to XAI tools, resources, and datasets, ensuring fairness and inclusivity in AI-driven art. Set up labs and workshops dedicated to inclusive design, where artists and technologists work together to develop XAI Arts tools that are accessible and empowering for all.

\rr{\subsubsection*{Accessible and Inclusive Design} Design XAI tools and platforms which incorporate actionable mechanisms to advance accessible, ethical, and inclusive AI practices. For example, building on studies of the use of generative AI tools by people with sensory impairments \cite{raman_exploring_2024}, and designing for artists with a range of technical literacies \cite{clemens_explaining_2023}. This aligns with alt.chi's values and our manifesto's potential to foster dialogue around ethical, human-centred AI through interdisciplinary collaboration.}

\subsubsection*{Accessibility Audits} Conduct regular explainability audits and build risk-registers of AI Arts tools and platforms to ensure they meet accessibility standards and address the needs of diverse user groups.

\subsubsection*{AI Literacy} Launch campaigns to educate the creative community about the challenges, benefits, and potential of XAI, focusing on transparency and ethical use of AI, and advocate for artist control of AI. Publish case studies of collaborations between artists and AI systems \rr{\cite{lewis_aixartist_2023, lewis_looking_2024}}, highlighting how AI can enhance creative practices without compromising artistic integrity, \rr{such as Holly Herndon and Mat Dryhurst's ``The Call'' exhibition at the Serpentine Gallery\footnote{\url{https://www.serpentinegalleries.org/whats-on/holly-herndon-mat-dryhurst-the-call/}} which foregrounded the curation of training data and training of AI models as a form of public AI literacy.}
Create training and AI literacy programs for artists to use AI tools effectively, with increased control of machine learning systems, and foster a sense of agency and confidence in their creative process through XAI approaches.

\subsection*{Valuing Artistic Practice for Explainability}
Current approaches to XAI are technocentric and typically originate from academic research labs with a focus on Human-Computer Interaction or Engineering, or are from major technology companies. 
This has resulted in XAI approaches driven by productivity, optimisation, and efficiency-oriented discourses. The Arts provide an alternative lens through which to drive XAI and offer the potential for radically alternative forms of explanation. We ask you to:

\subsubsection*{Aesthetic-Driven Research} Promote research that focuses on first-person accounts and explorations \rr{\cite{caramiaux_explorers_2022} of AI art's aesthetic and experiential aspects rather than purely technical explanations. For example, building on and engaging with longitudinal accounts of sustained artistic engagement with generative AI models in creative practice \cite{tecks_explainability_2024}.}

\subsubsection*{Artist-in-Residence Programs} Establish artist-in-residence programs \rr{\cite{serpentine_arts_technologies_future_2020}} within AI research labs to prioritise artistic exploration and control, and practice-led research on XAI. Organise residencies that bring artists and AI researchers together to create works that make AI models understandable through sound, visual, and interactive art. 

\subsubsection*{Collaborative workshops} Host workshops encouraging artists to explore new creative avenues with XAI, pushing the boundaries of traditional art forms and methods.

\subsubsection*{Demystifying AI} Develop educational materials, such as interactive installations, online platforms, videos, and articles, that explain the computational materiality of AI in accessible language and engaging formats.

\subsubsection*{AI Art Exhibitions} Host exhibitions showcasing artworks that explain AI concepts, accompanied by workshops and talks to engage the audience in understanding the underlying technology and its implications in art making and creative practices.

\subsection*{Hacking and Glitches}

The Arts offer practice-based approaches to challenging existing norms and discourses.
\rr{Prominent strategies involves employing the ideas of \textit{hacking} and \textit{glitch} as creative tools. These are commonly exploited by artists to exploring the sometimes unruly inner-workings of machines underneath the apparent order at the surface \cite{caramiaux_explorers_2022}. }
Hacking is often used in artistic practices to subvert, re-purpose, or reinterpret systems--be they technological, social or ideological. 
Similarly, glitch—traditionally regarded as a technological malfunction—is embraced as a deliberate aesthetic and conceptual choice.
\rr{By embracing error, whilst removing expectations of perfection for artists, also allows us to reveal the fragility and unpredictability of these systems.}
Such practices could be deployed to critically challenge and reflect on the use of AI and explanations of AI. We ask you to:

\subsubsection*{Hackathons} Regularly organise hackathons and workshops that encourage collaboration and experimentation with XAI, fostering a culture of innovation and creativity. \rr{For example, building on \citet{broad_using_2024}'s Hacker's Guide to using generative AI as an artistic material.}

\subsubsection*{Transparency and Surprise}
Promote projects that embrace glitches and imperfections in AI as artistic elements \rr{\cite{knight_artistic_2023}}, revealing the inner workings and boundaries of the technology through unexpected and playful outcomes instead of pragmatic explanations.

\subsection*{Openness}

Artistic practices can integrate knowledge across diverse disciplines. This interdisciplinary view offers combined thinking and sharing of methods to enable research in XAI that is at the same time technically robust, and also inclusive, intuitive, and engaging for broad audiences. Moreover, a cross-disciplinary view from the Arts ensures that insights and innovation are accessible to those without technical expertise. To increase interdisciplinarity in XAI research, especially greater inclusion of artists in XAI research, we ask you to:

\subsubsection*{Open Communities} Support community-driven projects that leverage XAI to create novel and impactful artworks, promoting a collaborative approach to AI art. \rr{Proactively reach out to underrepresented art groups to build more inclusive communities around AI and the Arts.}

\subsubsection*{Open Collaborations}
Facilitate collaborations between artists, technologists, and other disciplines to foster interdisciplinary innovation in AI art, and incorporate XAI into existing art practices. \rr{For example, engaging with and building on networks such as \citet{bryan-kinns_reducing_2024}'s international network on reducing bias in AI models for music.}

\subsubsection*{Open Access}
Support initiatives that provide open access to datasets and models, with transparency around where any data used was sourced from. Strive to reduce the homogenization of output and enable a wider range of creative experimentation and research on explanations and XAI. Encourage the creation and sharing of open-source AI projects (i.e., architectures, datasets, model weights, documentation), allowing artists and researchers to examine and modify the underlying code and data.

\subsubsection*{Open Tools and Datasets} Develop XAI tools, user interfaces, and collections of open-source datasets specifically designed for artists, allowing for personalisation, customisation and meaningful control over AI-generated outputs. \rr{Connecting to wider discourses on Future Art Ecosystems, openness requires developing more open and explainable Art x Public AI systems \cite{serpentine_arts_technologies_future_2020_4}.}

\section{Conclusion and Invitation}

By implementing the steps in this manifesto, we aim to demystify and humanize AI, making it more accessible, relatable, understandable, and explainable. The interdisciplinary collaborations sparked and catalysed by the XAIxArts manifesto will bridge the gap between technologists and artists, enriching both fields through mutual knowledge exchange.
We invite you to comment, amend, and change this manifesto in person at our alt.chi 2025 living manifesto session, and online through our living manifesto\footnote{XAIxArts Living Manifesto: \url{https://miro.com/app/board/uXjVL7tcbz8=/}}.
Your insights and contributions are essential in shaping a more transparent and harmonious future where AI and the Arts not only coexist but inform and enrich each other beyond current technocentric discourses. Let’s work together to keep this dynamic intersection of Explainable AI and the Arts vibrant, inclusive, and innovative.  

\section*{Accessibility Statement}
\rr{We use Miro to support co-creation of our living manifesto at alt.chi 2025 and previous XAIxArts workshops.} Miro is an accessible digital whiteboard with features including keyboard navigation, screen reader support, and zoom options, making it user-friendly for people with visual, motor, or cognitive impairments. \rr{Accessing Miro online also ensures that we are inclusive in our approach and engage with participants who are not able to engage in person with the manifesto - its intuitive design and assistive technology compatibility make it ideal for a iteratively co-created living manifesto in-person and online.}

\section*{Author disclosures}

This manifesto is original work; it was co-created by attendees (Terence Broad, Nick Bryan-Kinns, Francisco Castro, Jia-Rey Chang, Michael Clemens, Corey Ford, Jie Fu, Helen Kennedy, Makayla Lewis, Sophie Rollins, Karen Royer, Austin Tecks, Gabriel Vigliensoni, Elizabeth Wilson, and Shuoyang Zheng)vof the 2nd International Workshop on Explainable AI for the Arts (XAIxArts), part of ACM Creativity and Cognition conference 2024. Artificial Intelligence (ChatGPT and Grammar.ly) was utilized for copy-editing purposes only. 

\begin{acks}
Many thanks to the attendees of the ACM Creativity and Cognition 2024 conference who contributed their thoughts to revisions of the manifesto.
This work was supported in part by the RAI UK International Partnerships project ``Responsible AI international community to reduce bias in AI music generation and analysis'' [Responsible AI IP054 Grant Ref: EP/Y009800/1], the Bridging Responsible AI Divides programme with funds received from the Arts and Humanities Research Council [grant number AH/X007146/1], and the University of the Arts London Creative Computing Institute.
Shuoyang Zheng is a research student at the UKRI Centre for Doctoral Training in Artificial Intelligence and Music, supported by UK Research and Innovation [grant number EP/S022694/1].
\end{acks}

\section*{Appendix: Presentation at alt.chi 2025}

\rr{This alt.chi 2025 paper introduces a manifesto for Explainable Artificial Intelligence for the Arts (XAIxArts). We do this in the alt.chi forum to provoke rethinking of what it means to explain AI and at the same time to co-create with the alt.chi community version 2.0 of the XAIxArts manifesto.}

\rr{At the alt.chi session at CHI 2025 we transform the conventional conference presentation format into an interactive living manifesto experience by combining a World Caf\'e \cite{brown2021worldcafe} style discussion with a living manifesto wall.
We engage in small group discussions around caf\'e style tables (if available in the alt.chi venue or a nearby caf\'e space) to question each of our four manifesto themes described in Section \ref{sec:manifesto} one by one. Participants rotate between tables every 10 to 20 minutes as time permits in the alt.chi session. We will prepare living walls for the space on which participants can add comments using pens, post-it notes, and any other materials they have to have - large sheets of paper printed with each of the XAIxArts themes and calls to action.
We then invite harvesting of ideas and reflections and document them by commenting on and co-editing our living manifesto. The living manifesto will be hybrid at alt.chi 2025 to increase accessibility and engagement - participants can either co-edit the manifesto online or in person by writing directly on the pre-prepared living walls. 
The hybrid living manifesto remains available for comment throughout alt.chi 2025 and the physical parts are documented and moved online using open and accessible collaborative document editing tools\footnote{XAIxArts Living Manifesto: \url{https://miro.com/app/board/uXjVL7tcbz8=/}}. We aim to produce our updated XAIxArts manifesto 2.0 during the CHI conference so that it could be published online during CHI 2025. We will then keep our living manifesto alive online for ongoing editing and comment after the close of the conference at the XAIxArts website\footnote{\url{https://xaixarts.github.io/}}.}

\bibliographystyle{ACM-Reference-Format}


\end{document}